\documentclass[fleqn,usenatbib,useAMS]{mnras}

\usepackage{graphics}
\usepackage{graphicx}
\usepackage{dcolumn}
\usepackage{bm}
\usepackage{amsmath,amssymb}
\usepackage{hyperref}
\usepackage{epstopdf}
\usepackage[normalem]{ulem}
\hypersetup{
    colorlinks=true,
    linkcolor=blue,
    filecolor=magenta,      
    urlcolor=blue,
    citecolor=blue
}

\newcommand{\beq}{\begin{equation}}
\newcommand{\eeq}{\end{equation}}
\newcommand{\beqa}{\begin{eqnarray}}
\newcommand{\eeqa}{\end{eqnarray}}

\usepackage[T1]{fontenc}
\usepackage{ae,aecompl}

\usepackage{newtxtext,newtxmath}

\title[Measuring Individual Masses of Binary WDs with GWs]
{Measuring Individual Masses of Binary White Dwarfs with Space-based Gravitational-wave Interferometers}

\author[Anna Wolz, Kent Yagi, Nick Anderson, Andrew J. Taylor]{
Anna Wolz$^{1}$\thanks{Contact e-mail: \href{mailto:amw9cj@virginia.edu}{amw9cj@virginia.edu}}, 
Kent Yagi$^{1}$\thanks{Contact e-mail: \href{mailto:ky5t@virginia.edu}{ky5t@virginia.edu}}, 
Nick Anderson$^{2}$\thanks{Contact e-mail: \href{mailto:nwa2ub@virginia.edu}{nwa2ub@virginia.edu}}, 
Andrew J. Taylor$^{2}$\thanks{Contact e-mail: \href{mailto:ajt9gx@virginia.edu}{ajt9gx@virginia.edu}}
\\
$^{1}$Department of Physics, University of Virginia, Charlottesville, VA\\
$^{2}$Department of Astronomy, University of Virginia, Charlottesville, VA
}

\date{Accepted XXX. Received YYY; in original form ZZZ}

\pubyear{2020}

\begin{document}
\label{firstpage}
\pagerange{\pageref{firstpage}--\pageref{lastpage}}
\maketitle

\begin{abstract}
Unlike gravitational waves from merging black holes and neutron stars that chirp significantly over the observational period of ground-based detectors, gravitational waves from binary white dwarfs are almost monochromatic. This makes it extremely challenging to measure their individual masses. Here, we take a novel approach of using finite-size effects and applying certain universal relations to measure individual masses of binary white dwarfs using LISA. We found quasi-universal relations among the mass, moment of inertia, and tidal deformability of a white dwarf that do not depend sensitively on the white dwarf composition. These relations allow us to rewrite the moments of inertia and tidal deformabilities in the waveform in terms of the masses. We then carried out a Fisher analysis to estimate how accurately one can measure the individual masses from the chirp mass and finite-size measurements. We found that the individual white dwarf masses can be measured with LISA for a 4-year observation if the initial frequency is high enough ($\sim 0.02$Hz) and either the binary separation is small ($\sim 1$kpc) or the masses are relatively large $(m \gtrsim 0.8M_\odot)$.
This opens a new possibility of measuring individual masses of binary white dwarfs with space-based interferometers.

\end{abstract}




\def\lesssim{\mathrel{\hbox{\rlap{\hbox{\lower5pt\hbox{$\sim$}}}\hbox{$<$}}}}
\def\gtrsim{\mathrel{\hbox{\rlap{\hbox{\lower5pt\hbox{$\sim$}}}\hbox{$>$}}}}

\section{Introduction}

Recent direct detection of gravitational waves (GWs) by the LIGO/Virgo Collaborations~\citep{Abbott:2016blz,TheLIGOScientific:2017qsa,LIGOScientific:2018mvr} ushered in a new field of GW astronomy. While ground-based detectors will be upgraded to third-generation ones, there are plans to launch space-based detectors that will have sensitivities in a frequency band lower than ground-based ones. The Laser Interferometer Space Antenna (LISA)~\citep{2017arXiv170200786A} is expected to be launched in 2034 with its best sensitivity around $0.01$Hz (other similar plans exist, such as TianQin~\citep{Luo:2015ght}).
The Deci-hertz Interferometer Gravitational wave Observatory (DECIGO)~\citep{setoDECIGO,2020arXiv200613545K} bridges the gap between ground-based detectors and LISA/TianQin. 

Binary white dwarfs (WDs) are promising targets for LISA. For example, recent simulations by~\citet{Lamberts:2019nyk} show that roughly 12,000 binary WDs will be individually resolved for an observation period of 4 years\footnote{Using different input models including the binary evolution model and the spatial distribution of the galaxy predicts a slightly different number of detections, see e.g.~\citep{2018MNRAS.480..302K}.}. Unlike binary neutron stars and stellar-mass binary black holes that are targets for ground-based detectors, GW signals from binary WDs for LISA are almost monochromatic with a small amount of chirp in frequency. This means that LISA can measure the frequency $f$ and possibly its time derivative $\dot f$~\citep{2012A&A...544A.153S}, from which one can extract the chirp mass, 
while it is extremely difficult to measure the mass ratio, or in turn the individual masses. Meanwhile, if the frequency is high enough ($f \gtrsim 0.01$Hz), there can be enough chirp to also measure $\ddot f$, which can be used to measure finite-size effects. These depend on the internal structure of a WD and are caused by the tidal field of a companion WD and its own rotation that we assume to be synchronous\footnote{Synchronous rotation assumes a large amount of tidal friction such that the synchronization time is much shorter than the orbital decay time. This is a key difference from binary neutron stars in which tidal frictions are thought to be negligible.}~\citep{Shah:2014oea}. 

In this \emph{letter}, we propose a novel way of measuring the individual masses of binary WDs through the finite-size effects. Such effects are characterized by the moment of inertia and tidal deformability (or tidal Love number). 
Relations between (properly-normalized) moment of inertia and tidal deformability (I-Love relation) are known to be quasi-universal (i.e.~insensitive to the stellar internal structure) for neutron stars~\citep{I-Love-Q-Science,Yagi:2013awa,Yagi:2016bkt}, which were also extended to cold WDs~\citep{ilq_wd,Taylor:2019hle}. Moreover, we show that universal relations also exist between the moment of inertia (normalized by mass cubed in the geometric units) and the mass (I-M relation). Adopting these two different universal relations, one can express the finite-size effects of binary WDs in gravitational waveforms in terms of the individual masses. Thus, a measurement of the finite-size effects will lead to a measurement of the individual masses. 

We carry out a Fisher analysis to demonstrate this idea and reveal the measurability of the individual masses of binary WDs with GWs. We only consider WDs with masses larger than 0.3$M_\odot$ for which the finite-temperature effect is expected to be small.
We use the results in~\cite{Benacquista:2011gh} and~\citet{Shah:2014oea} for the finite-size effects in the waveform but we correct the tidal deformability part, following the analysis in binary neutron star waveforms~\citep{flanagan-hinderer-love}.

We work in the geometric units of $c=G=1$ throughout, which means that the mass, length and time all have the same dimension. One can easily recover the true, physical dimension using the conversion $1M_\odot = 1.5\mathrm{km} = 4.9\times 10^{-6}$s.

\section{Gravitational Waveform}
 
We begin by presenting gravitational waveforms for binary WDs with finite-size effects. We consider two such effects, (i) rotation effects and (ii) tidal effects. We assume that the 
WD rotations are synchronized with the orbital motion. In such a case, WD rotations enter in the waveform formally at second post-Newtonian (PN) order\footnote{PN order counts the power of velocity squared relative to the leading, Newtonian order for the point-particle motion.} through the moment of inertia $I$~\citep{Benacquista:2011gh}. On the other hand, the tidal effect enters through the tidally-induced quadrupole moment $Q_{ij}$, which is characterized by a linear response function called tidal deformability $\lambda$ as $Q_{ij} = - \lambda \mathcal E_{ij}$~\citep{hinderer-love},
where $\mathcal E_{ij}$ is the external quadrupolar tidal field.
$\lambda$ is related to the tidal Love number $k_2$ as $k_2 = (3/2) \lambda/R^5$~\citep{hinderer-love}, where $R$ is the WD radius. $Q_{ij}$ changes the binding energy of a binary, Kepler's law, and the GW luminosity from the point-particle case. It is well known that such a tidal effect formally enters at 5PN order~\citep{flanagan-hinderer-love}. 

The evolution of the orbital angular velocity $\omega = \pi f$ for the GW frequency $f$ is given by\footnote{The last term corrects Eq.~(10) of~\citet{Benacquista:2011gh} for the contribution from the tidally-induced quadrupole moment, which was derived by assuming that $Q$ was independent of $\omega$.}
\begin{equation}
\label{eq:omega_dot}
    \dot{\omega} = \dot{\omega}_{0\textrm{PN}}(1+\Delta_{1\textrm{PN}}x+\Delta_{\textrm{I}}x^2+\Delta_{\Lambda}x^5)\,,
\end{equation}
with\footnote{Notice that the definition of $\Delta_\textrm{I}$ is slightly different from that in~\citet{Benacquista:2011gh}.}
\allowdisplaybreaks
\begin{align}
    \dot{\omega}_{0\textrm{PN}}=&\frac{96}{5} M_c^{5/3}\omega^{11/3}\,, \\
    \Delta_{1\textrm{PN}}=&-\Big(\frac{743}{336}+\frac{11}{4}\eta\Big)\,, \\
       \Delta_{\textrm{I}}=&\frac{3}{\eta}(X_1^3\bar I_1+X_2^3\bar I_2)\,, \\
\label{tidal}
    \Delta_{\Lambda}=&\frac{39}{8}\tilde\Lambda\,.
\end{align}
Here $\eta = m_1 m_2/M^2$ is the symmetric mass ratio
with individual WD masses $m_A$ (with $m_1 \geq m_2$) and $M = m_1 + m_2$ representing the total mass, $M_c = M \eta^{3/5}$ is the chirp mass, $X_A = m_A/M$,  $x = (M \omega)^{2/3}$ corresponding to the relative velocity squared of the WDs in a binary, $\bar I_A = I_A/m_A^3$ is the dimensionless moment of inertia of the $A$th body\footnote{We normalize the moment of inertia in this way since the universality depends sensitively on how one normalizes each quantity~\citep{Yagi:2016bkt}.}, and~\citep{Wade:2014vqa}
\begin{align}
    \tilde\Lambda & =\frac{8}{13}\Big[(1+7\eta-31\eta^2)(\Lambda_1+\Lambda_2) \nonumber \\
    & \quad +\sqrt{1-4\eta}(1+9\eta-11\eta^2)(\Lambda_1-\Lambda_2)\Big]\,,
\end{align}
with $\Lambda_A = \lambda_A/m_A^5$ corresponding to the dimensionless tidal deformability. 
The first term of Eq.~\eqref{eq:omega_dot} is the leading point-particle term and the second term is the 1PN point-particle contribution~\citep{arun35PN}, while the third and fourth terms are due to the moment of inertia~\citep{Benacquista:2011gh} and the tidal deformability~\citep{flanagan-hinderer-love}, respectively.
Similar to Eq.~\eqref{eq:omega_dot}, the second time derivative of $\omega$ is given by
\begin{align}
    \ddot{\omega}=\frac{11}{3}\frac{\dot{\omega}_{0\textrm{PN}}^2}{\omega}\Big(1+\frac{24}{11}\Delta_{1\textrm{PN}}x+\frac{26}{11}\Delta_{\textrm{I}}x^2+\frac{32}{11}\Delta_\Lambda x^5\Big)\,.
\end{align}

The gravitational waveform $h(t)$ consists of a linear combination of the plus mode $h_+(t)$ and the cross mode $h_\times (t)$:
\begin{equation}
h(t) = F_+ h_+(t) + F_\times h_\times(t)\,,
\end{equation}
where $F_+$ and $F_\times$ are the beam-pattern functions for the two modes~\citep{cutler1998} that depend on the sky position of the binary and the polarization angle. Performing the sky-averaging, the waveform becomes
\begin{equation}
h(t) = A\cos \phi(t)\,,
\end{equation}
with the amplitude
\begin{align}
    A=\frac{8M_c}{5D}(\pi M_c f_0)^{2/3}\,,
\end{align}
and the phase~\citep{Shah:2014oea}
\begin{equation}
\label{phi}
    \phi(t)= \phi_0+2\omega_0\delta t+\dot{\omega}_0 \delta t^2+\frac{1}{3}\ddot{\omega}_0\delta t^3\,,
\end{equation}
where $D$ is the luminosity distance, the subscript 0 represents the quantity evaluated at the initial observation time $t_0$, and $\delta t=t-t_0$. 

\begin{figure}
\centering\includegraphics[width=7.5cm]{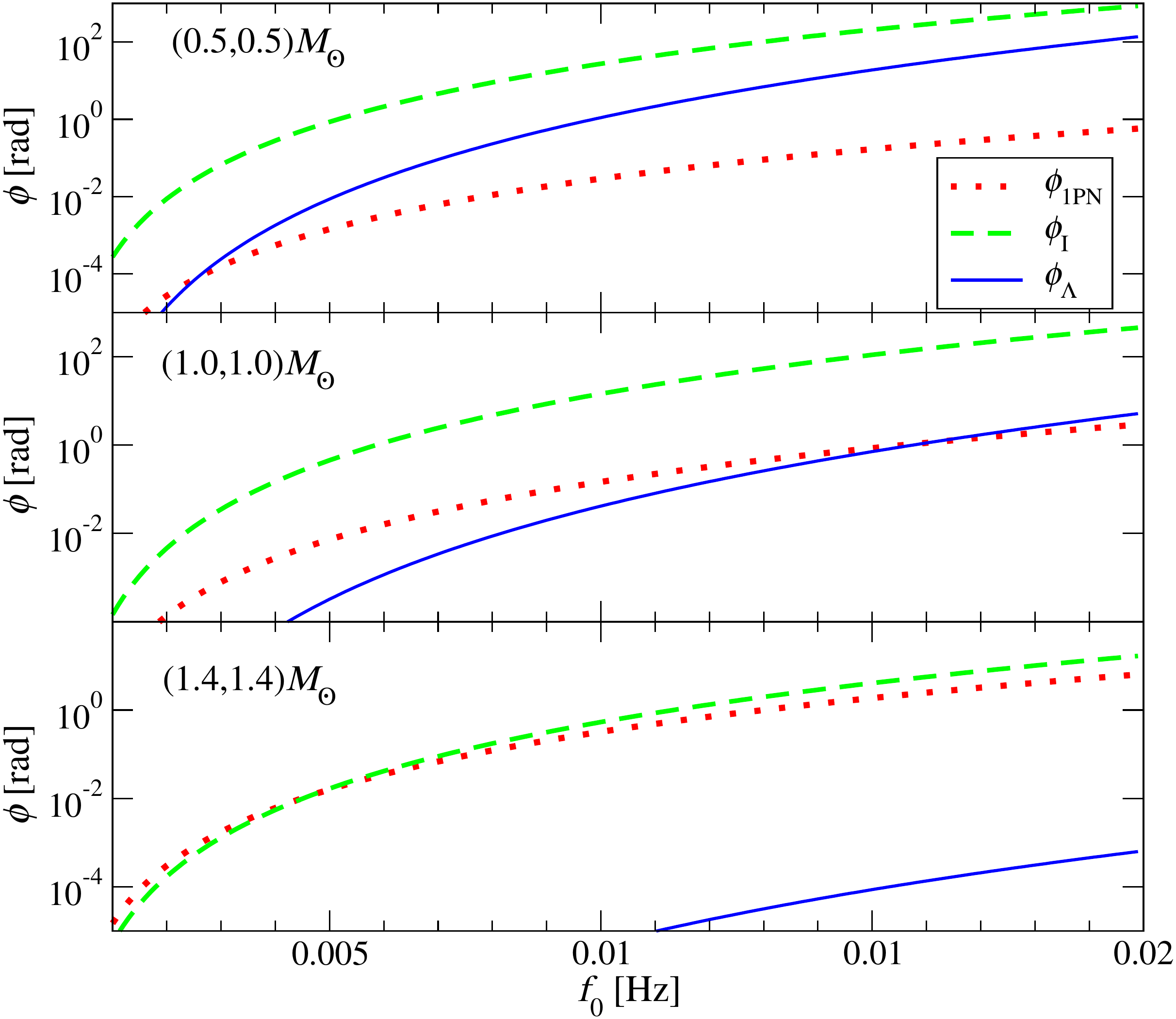}
\caption{Effects on the GW phase (accumulated over an observation period of 4 years) due to the 1PN effect ($\phi_{\textrm{1PN}}$), moment of inertia effect ($\phi_{\textrm{I}}$), and tidal deformability effect ($\phi_{\Lambda}$) as a function of the initial frequency $f_0$ for binary WDs with three different mass combinations. 
}\label{phaseeffects}
\end{figure}

Let us compare the phase contributions due to the 1PN ($\phi_\mathrm{1PN}$), moment of inertia ($\phi_\mathrm{I}$), and tidal deformability ($\phi_\Lambda$) terms. 
Figure~\ref{phaseeffects} compares these contributions of the phase accumulated over an observation period of 4 years as a function of the initial GW frequency $f_0$ for various mass combinations. 
Given two masses $m_1,m_2$ in a binary WD system, we obtained $I$ and $\Lambda$ from the data produced in~\citep{Taylor:2019hle}\footnote{We chose the atomic number of the WD equation of state as $Z=0$ though these quantities are insensitive to the choice of $Z$.}. Observe that the tidal effect can be close to 50\% of the moment of inertia effect. Observe also that the 1PN effect can be comparable to the moment of inertia effect when the masses are relatively large. 
For these reasons, we will include all of these effects in our analysis. 

\section{Universal Relations}

We wish to determine individual masses of a binary WD with GWs. To do so, we need to relate the moment of inertia and tidal deformability of a WD to its mass. Luckily, such relations exist for WDs since the equation of state for WDs is more or less known. These relations are similar to the universal relations for neutron stars discussed e.g. in~\citet{I-Love-Q-Science,Yagi:2013awa,Yagi:2016bkt}. We focus on studying cold, slowly-rotating WDs. The former assumption is a good approximation for WDs with relatively large masses~\citep{2007MNRAS.382..779P}. 
For the latter assumption, we checked that higher order corrections are smaller than $\sim 10$\% for synchronized WDs when $f \leq 0.2$Hz. 

\begin{figure}
\centering\includegraphics[width=7.5cm]{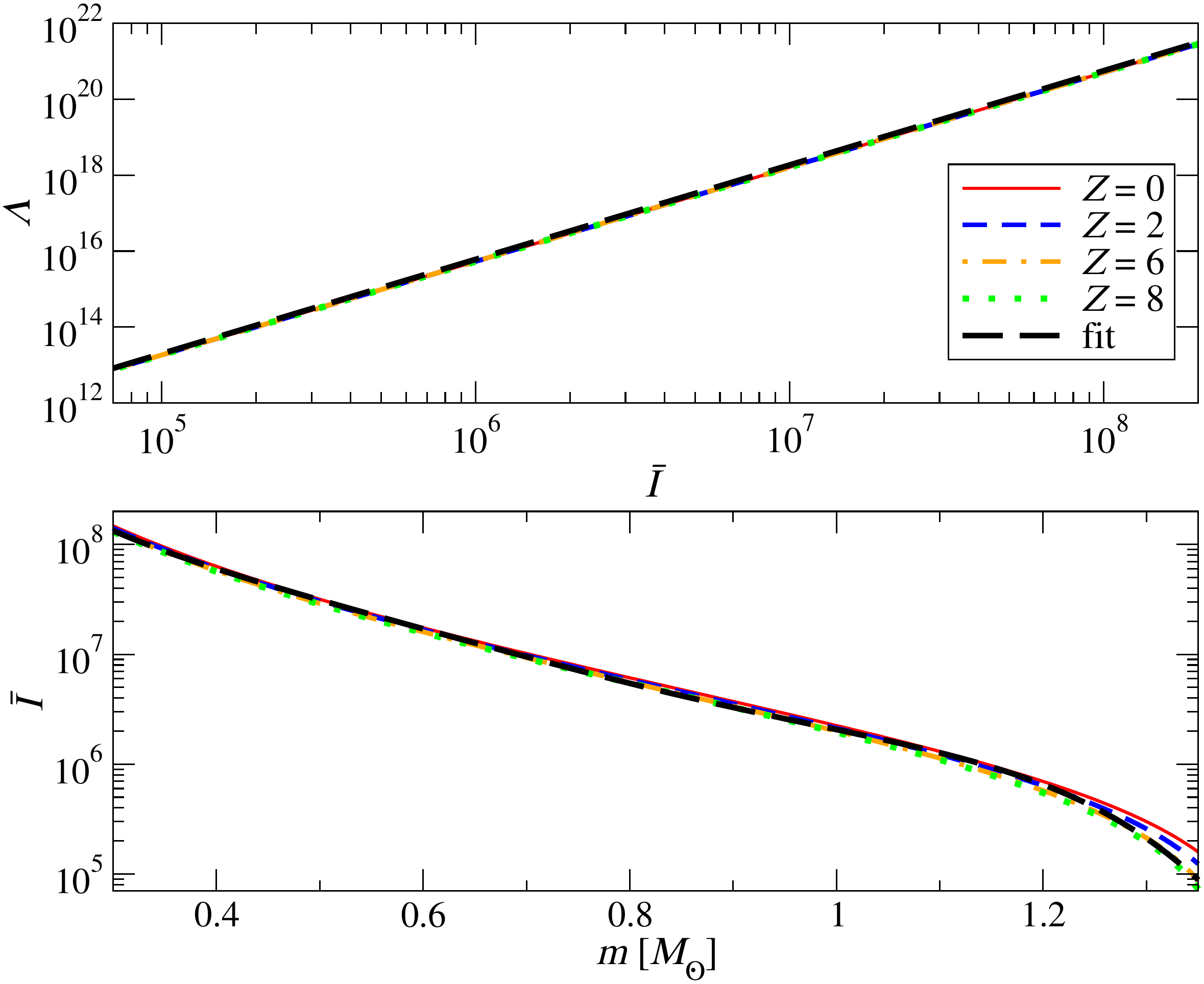}
\caption{
Universal relations between the dimensionless tidal deformability and moment of inertia (top), universal to an error of 
0.1\% and between the dimensionless moment of inertia and the mass (bottom), universal to an error of 5\%.
We show results for 4 different values of the atomic number $Z$ based on the data sets in~\citet{Taylor:2019hle}. We also present the fit given by Eqs.~\eqref{eq:I-Love_fit} and~\eqref{eq:I-M_fit}. 
}\label{universal}
\end{figure}

We adopt two universal relations. One is between the moment of inertia and the tidal deformability (the so-called I-Love relation)~\citep{ilq_wd,Taylor:2019hle}.
We present the relation in the top panel of Fig.~\ref{universal} with various atomic numbers $Z$ for the WD equation of state that assumes degenerate (but arbitrarily relativistic) electron pressure with the electrostatic correction. Observe that the relation is insensitive to the choice of $Z$\footnote{In reality, one expects to have He at low masses, a C/O mixture at intermediate masses, and O/Ne/Mg at the highest masses. This means that we do not expect each $Z$ to be possible at each mass, and thus the $Z$-variation presented in Fig.~\ref{universal} is conservative.}. One can fit the relation as
\begin{align}
 \label{eq:I-Love_fit}
 \ln\Lambda=2.02942+2.48377 \ln\bar I\,,
 \end{align}
which we also present in the figure.
The second universal relation that we use is the one between the dimensionless moment of inertia and the WD mass, as shown in the bottom panel of Fig.~\ref{universal}. We constructed a fit of the form 
\begin{align}
\label{eq:I-M_fit}
    \ln \bar I& =24.7995 - 39.0476 m_{1M_\odot} + 95.9545 m_{1M_\odot}^2 \nonumber \\
    & \quad- 138.625 m_{1M_\odot}^3+ 98.8597 m_{1M_\odot}^4 - 27.4000 m_{1M_\odot}^5\,,
\end{align}
with $m_{1M_\odot}=m/1M_\odot$, where $m$ is the mass of a WD. We use these two relations to turn $\bar I_A$ and $\Lambda_A$ in the waveform to $m_A$.

\section{Parameter Estimation}

We estimate the error on the GW parameters using Fisher information analysis~\citep{cutler1998,2012A&A...544A.153S,Shah:2014oea}. The Fisher matrix provides information about the measurement errors of GW parameters, provided we have accurate gravitational waveform templates, assuming that the detector noise is stationary and Gaussian, and that the signal-to-noise ratio (SNR) is high ($\gg 10$).

The 1-$\sigma$ measurement uncertainty of parameters $\theta^i$ is given by
\begin{equation}
\Delta \theta_i = \sqrt{(\Gamma^{-1}){}_{ii}}\,.
\end{equation}
Here $\Gamma_{ij}$ is the Fisher matrix defined by
\begin{equation}
\Gamma_{ij} = \left( \frac{\partial h}{\partial \theta^i} \bigg|\frac{\partial h}{\partial \theta^j} \right)\,,
\end{equation}
where the inner product is defined by
\begin{equation}\label{eq:inner}
(a|b) = 4 \int^\infty_0 \frac{\tilde a^*(f) \tilde b(f)}{S_n(f)}df \approx \frac{2}{S_n(f_0)} \int^T_0 a(t) b(t) dt\,,
\end{equation}
with $T$ being the observation time, a tilde denoting the Fourier component and * representing complex conjugate.  $S_n$ is the spectral noise density and the one for LISA is given in~\citet{Cornish:2018dyw} as $\sqrt{S_n(0.02\mathrm{Hz})} = 1.43\times10^{-20} \, \mathrm{Hz}^{-1/2}$  while that for DECIGO is given in~\citet{Yagi:2009zz,Yagi:2011wg} as $\sqrt{S_n(0.02\mathrm{Hz})} = 6.46\times10^{-23} \, \mathrm{Hz}^{-1/2}$~\footnote{For DECIGO, $S_n$ is given in Eq.~(1) of~\citet{Yagi:2009zz} where the (updated) instrumental spectral noise density is in Eq.~(5) of~\citet{Yagi:2011wg}. We consider 4 triangular interferometers corresponding to 8 independent L-shaped interferometers.}. In the second equality of Eq.~\eqref{eq:inner}, we made an approximation that the signal for binary WDs is almost monochromatic so that $S_n(f) \approx S_n(f_0)$ and used the Parseval's theorem to turn the Fourier integral into a time integral~\citep{cutler1998,2012A&A...544A.153S}.

\begin{figure*}
\centering\includegraphics[width=0.43\textwidth]{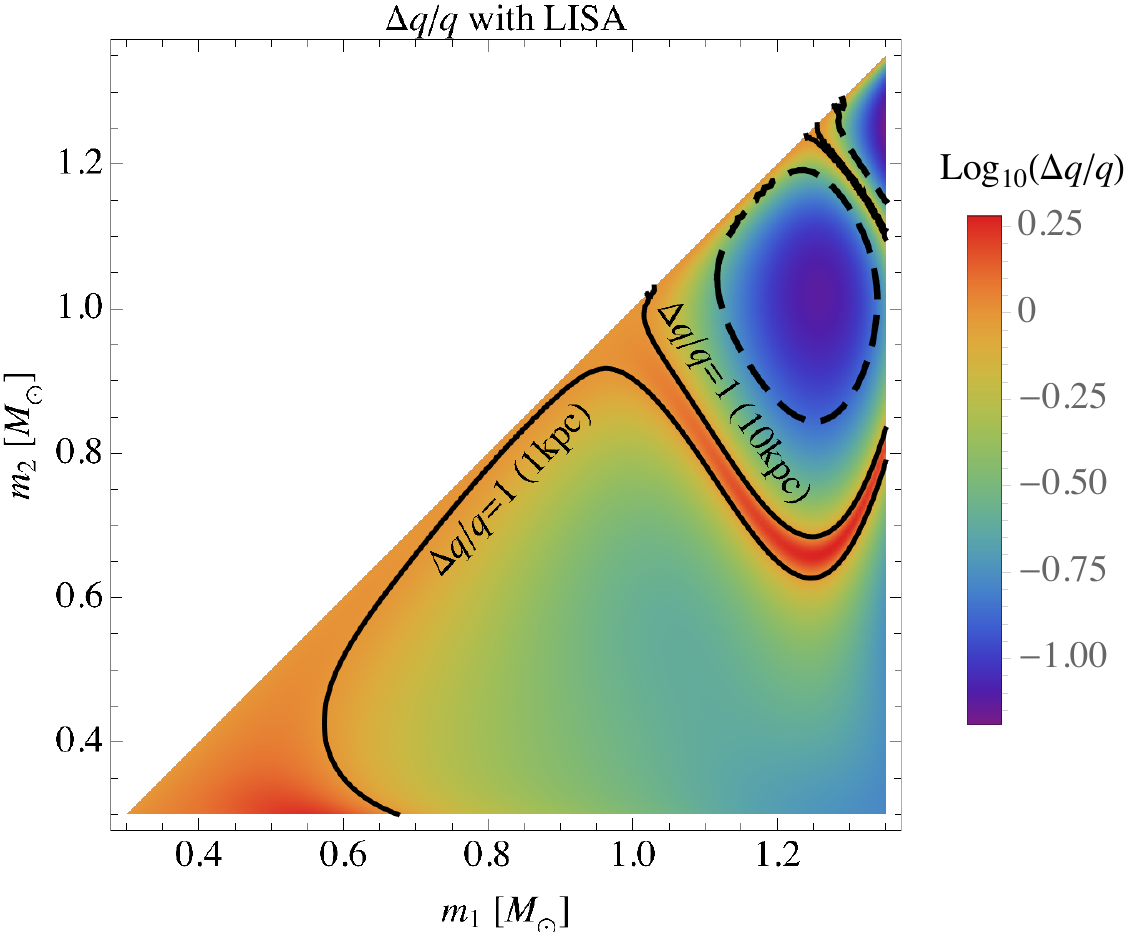} 
\centering\includegraphics[width=.43\textwidth]{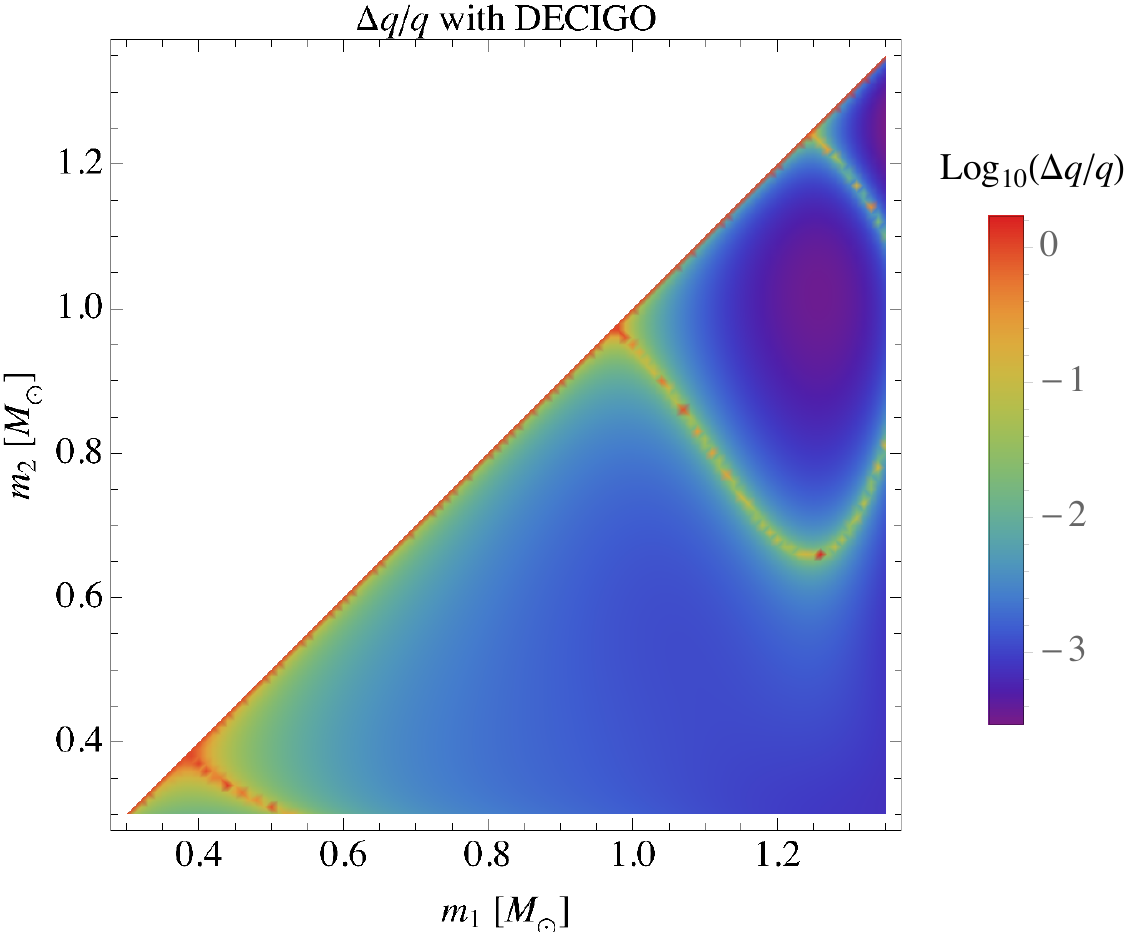} 
\caption{(Left) Fractional measurement error of the mass ratio $q$ at initial observation frequency $f_0=0.02$Hz and distance $D=1$kpc using LISA. The color coding indicates $\log_{10}(\Delta q/q)$ (e.g. $-1.00$ means $\Delta q/q = 10^{-1}$). The solid black contours represent the mass ratio measurability threshold of $\Delta q/q=1$. The SNRs vary with the masses and the minimum SNR in our analysis is $3.17 \times 10^2$. For comparison, we also show the threshold for binary WDs at 10kpc in black dashed contours. 
(Right) Similar to Fig.~\ref{fig:qmeas020log} but with DECIGO. 
The minimum SNR in this case is $7.02 \times 10^4$.
}
\label{fig:qmeas020log}
\end{figure*}

We consider two sets of parameters. The first set is 
\begin{equation}
\label{eq:parameter1}
\theta^i = (M_c, q, f_0, \phi_0, A)\,,
\end{equation}
where $q = m_2/m_1$ is the mass ratio\footnote{Since the Fisher matrix is given in terms of the derivative of each parameter with respect to the waveform, the measurability of $q$ depends heavily on $\partial \phi/\partial q$, in particular $\partial \dot f/\partial q$, as we will discuss later.}.
After converting $\bar I_A$ and $\Lambda_A$ in the waveform into $m_A$ through the universal relations, we further convert $m_A$ to $M_c$ and $q$. We also impose the prior that $q \leq 1$. For simplicity, we impose a Gaussian prior~\citep{cutlerflanagan} on $q$ such that the measurement uncertainty of $q$ is now corrected to~\citep{Carson:2019kkh}
\begin{equation}
\Delta \theta_i = \sqrt{(\tilde \Gamma^{-1}){}_{ii}}\,, \quad \tilde \Gamma_{ij} = \Gamma_{ij}+\frac{1}{\sigma_{\theta^i}^2} \delta_{ij}\,,
\end{equation}
with $\sigma_q = 1$\footnote{We do not impose prior on other parameters, which effectively corresponds to setting $\sigma_{\theta^i} = \infty$  for parameters other than $q$. One could further impose the prior on $\phi_0$ to restrict it to in the range $[0,2\pi]$, though we checked that such a prior does not change the results shown.}.
The second set of parameters that we consider is 
\begin{equation}
\label{eq:parameter2}
\theta^i = (m_1,m_2, f_0, \phi_0, A)\,,
\end{equation}
so that we can directly compute the measurement uncertainty of the individual masses in the binary. We can convert $M_c$ and $q$ in the waveform to $m_A$ using 
\begin{align}\label{m1}
    m_1(M_c,q)=&M_c q^{-3/5} (q+1)^{1/5}\,, \\
\label{m2}
    m_2(M_c,q)=&M_c q^{2/5}(q+1)^{1/5}\,.
\end{align}
We take $f_0=0.02$Hz, $\phi_0=3.666$ rad, and $D=1$kpc (or 10kpc) as fiducial values unless otherwise stated, and vary $(M_c,q)$ or $(m_1,m_2)$. We mainly consider the observation time of $T=4$ yr.

\section{Results}

We present the main results by only considering WDs with masses larger than $0.3 M_\odot$ in order to keep the cold WD approximation.
We begin by studying the results for the first parameter set in Eq.~\eqref{eq:parameter1}. 
The left panel of Fig.~\ref{fig:qmeas020log} presents the fractional measurement error of $q$ as a function of $m_1$ and $m_2$ with LISA (the measurability of $M_c$ is much better than that for $q$). The detection threshold of $\Delta q/q=1$ is shown by the black solid contours for $D=1$kpc. Observe that the mass ratio can be measured in most of the $m_1$--$m_2$ parameter region. Some of the region where the mass ratio cannot be measured arises partially from $\partial \dot f/\partial q = 0$. It also arises from
the degeneracy between $q$ and other parameters, in particular $M_c$. Although we chose the fiducial distance of $D=1$kpc in Fig.~\ref{fig:qmeas020log}, $q$ is still measurable for $D=10$kpc when the masses are relatively large $(m\geq0.8M_\odot)$. We also found that the measurability of $q$ with a 2-year observation and $D=1$kpc is similar to the result with a 4-year observation and $D=10$kpc\footnote{The shorter the duration is, the less chirp the binary becomes, which increases the amount of correlation among parameters. That is why the measurability of $q$ gets much worse than a simple $\sqrt{T}$ scaling.}. On the other hand, the measurability depends sensitively on the initial frequency $f_0$. For example, $q$ becomes immeasurable when $f_0 = 0.01$Hz. 

The situation improves further if we use DECIGO, as shown in the right panel of Fig.~\ref{fig:qmeas020log}. The measurement accuracy of $q$ increases from the LISA case by two orders of magnitude or more in most of the parameter space. However, given that there is a variation in the universal relations that can be as large as $5\%$ (plus rotation correction that can be higher than 5\% for low-mass WDs), there is a systematic error on the mass ratio with a similar amount. We checked that even with DECIGO, it is crucial to include the finite-size effects to measure individual masses for binary WDs considered here.

\begin{figure}
\centering\includegraphics[width=.45\textwidth]{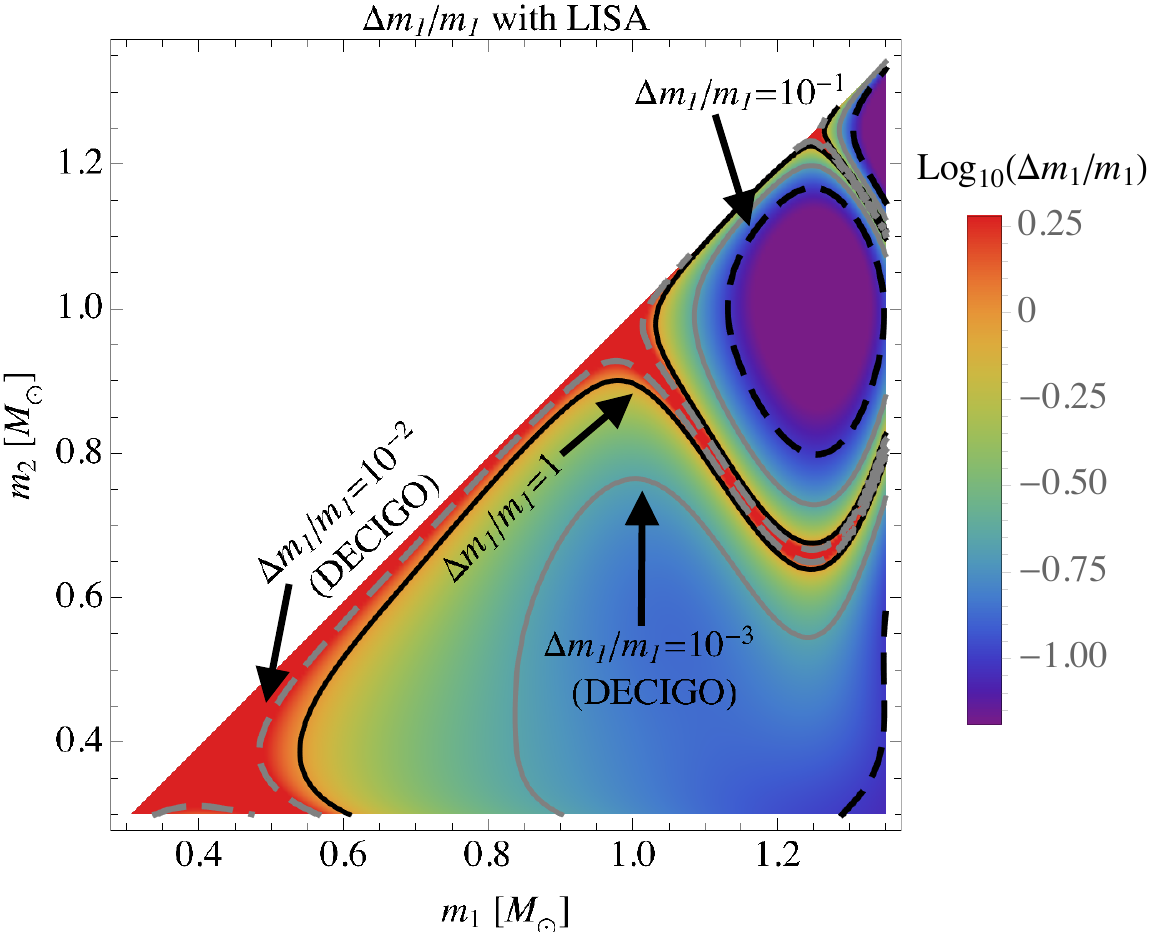} 
\caption{
Similar to Fig.~\ref{fig:qmeas020log} but for the fractional measurement error of primary mass $m_1$ for binaries at 1kpc.
We have made the color coding to be the same as Fig.~\ref{fig:qmeas020log} and the log fractional errors higher than $0.25$ are all shown as red. We present contours for $\Delta m_1/m_1 = 1$ (black solid) and $\Delta m_1/m_1 = 10^{-1}$ (black dashed) using LISA. For comparison, we also show contours for $\Delta m_1/m_1 = 10^{-2}$ (grey dashed) and $\Delta m_1/m_1 = 10^{-3}$ (grey solid) using DECIGO.
}\label{fig:m13}
\end{figure}

Let us next look at the results for the second parameter set in Eq.~\eqref{eq:parameter2}. Figure~\ref{fig:m13} presents the fractional measurability of $m_1$ (the one for $m_2$ is almost identical). Observe that $\Delta m_1/m_1$ is very similar to $\Delta q/q$ in most of the region, including the threshold contours of $\Delta m_1/m_1=1$. We found that the best fractional error one can obtain with these values is 1.7\%  with masses $(m_1,m_2)=(1.4,1.26) M_{\odot}$\footnote{This is roughly consistent with \citet{Shah:2014oea}, who found that $\ddot f$ (and in turn the moment of inertia) can be measured with a fractional error of 18\% for $m_1 \sim m_2 \sim 1 M_\odot$ at $f_0 \sim 0.02$Hz and $D \sim 10$kpc. If we rescale this to a binary at $D=1$kpc, the error on the moment of inertia becomes $\sim 1.8\%$.}, though again, this is limited by systematic errors from the variation in the universal relations and rotation corrections.
These findings show that LISA may be able to measure individual masses of binary WDs if the initial frequency is sufficiently high ($\sim 0.02$Hz). We also present in Fig.~\ref{fig:m13} a few contours using DECIGO. Similar to the $q$ measurability case, the overall trend of contours are similar to the LISA case, but the measurement accuracy can improve by more than two orders of magnitude.

\section{Conclusions}

We studied the possibility of measuring the individual masses in a binary WD using space-based GW detectors like LISA and DECIGO. We use the measurement of the finite-size effects of WDs that are characterized by the moment of inertia and tidal deformability. We then convert this to information on the mass using the universal relations between these tidal parameters (I-Love relation), and between the moment of inertia and the mass (I-M relation). We found that the individual masses can be measured with LISA for most of the mass combinations that we studied for our fiducial choice of $f_0=0.02$Hz and $D=1$kpc. If the WD masses are large enough $(m\gtrsim 0.8M_\odot)$, the individual masses can still be measured even if the binary is at 10kpc. 
Although a binary WD with large $f_0$ and either large masses or smaller distance (that is ideal to measure the individual WD masses with LISA) might be rare~\citep{Korol:2017qcx}, the result presented here opens a novel, interesting possibility. Furthermore, the chance of measuring individual WD masses increases significantly with DECIGO.

Various avenues exist for future work. For example, it would be interesting to use the population of simulated binary WDs in~\citet{Lamberts:2019nyk} and carry out a Bayesian Markov-chain Monte-Carlo analysis in the frequency domain~\citep{Cornish:2007if} to confirm the results presented in this letter. It would also be important to fold in the uncertainties in the universal relations to the measurability of the individual masses~\citep{Chatziioannou:2018vzf,Carson:2019rjx}. Furthermore, one may extend the current analysis by studying the case for asynchronized binaries~\citep{1981A&A....99..126H}. Finally, \cite{2019arXiv191001063K} pointed out that DECIGO may detect 50-20000 (extragalactic) binary WD \emph{mergers}. Since the chirp effect is larger for such binaries than those at $0.02$Hz, one can study whether DECIGO can measure both individual masses and finite-size effects of binary WDs independently.

\section*{Acknowledgements}
We thank Phil Arras and Tyson Littenberg for valuable comments. 
K.Y. acknowledges support from NSF Award PHY-1806776, NASA Grant 80NSSC20K0523, a Sloan Foundation Research Fellowship and the Owens Family Foundation. 
K.Y. would like to also acknowledge support by the COST Action GWverse CA16104 and JSPS KAKENHI Grants No. JP17H06358.

\section*{Data Availability}

The data underlying this article will be shared on reasonable request to the corresponding author.

\bibliographystyle{mnras}
\bibliography{ref}

\bsp	
\label{lastpage}
\end{document}